\documentclass[aps,twocolumn,prb,showpacs,superscriptaddress]{revtex4}
\usepackage{amssymb}
\usepackage{epsfig}
\usepackage{epsf}

\def\avg#1{\langle#1\rangle}

\def\be{\begin{equation}} \def\ee{\end{equation}}
\def\bea{\begin{eqnarray}} \def\eea{\end{eqnarray}}

\def\nn{\nonumber}
\def\pp{\parallel}

\begin{document}
\title{Unconventional metamagnetic electron states in orbital band 
systems}
\author{Wei-Cheng Lee}
\affiliation{Department of Physics, University of California, San Diego,
CA 92093}
\author{Congjun Wu}
\affiliation{Department of Physics, University of California, San Diego,
CA 92093}

\begin{abstract}
We extend the study of the Fermi surface instability of the Pomeranchuk type 
into systems with orbital band structures, which are common
features in transition metal oxides.
Band hybridization significantly shifts the spectra weight of the
Landau interactions from the conventional $s$-wave channel to
unconventional non-$s$-wave channels,
which results in anisotropic (nematic) Fermi 
surface distortions even with ordinary interactions in solids.
The Ginzburg-Landau free energy is constructed by
coupling the charge-nematic, spin-nematic and ferromagnetic order
parameters together, which shows that nematic electron states
can be induced by metamagnetism.
The connection between this mechanism to the anisotropic metamagnetc
states observed in Sr$_3$Ru$_2$O$_7$ at high magnetic fields is studied
in a multi-band Hubbard model with the hybridized
quasi-one dimensional $d_{xz}$ and $d_{yz}$-bands.
\end{abstract}
\maketitle

\section{Introduction}
Pomeranchuk instabilities  are a large class of Fermi surface 
instabilities in both density and spin channels, each of which 
can be further decomposed into different partial wave
channels \cite{pomeranchuk1959}.
This class of instabilities 
in the non-$s$-wave density channel result in uniform but anisotropic
(nematic) electron liquid states \cite{oganesyan2001}, which have 
recently attracted a great deal of attention in recent years
\cite{oganesyan2001,barci2003,barci2008,lawler2006,Nilsson2006,quintanilla2006,halboth2000,dellanna2006,kee2003,khakvine04,
yamase00,yamase01,yamase05,yamase07,kivelson2003,varma2005,kee2005,honerkamp2005,wolflel2007, lamas2008}.
In particular, these instabilities have been studied in the context of
doped Mott insulators~\cite{kivelson1998}, high T$_c$ 
materials \cite{kivelson1998,kivelson2003}, and quantum Hall systems in 
nearly half-filled Landau levels \cite{fradkin99,fradkin2000}.
Experimental evidence has also been found in ultra-high mobility 
two-dimensional electron gases  
and quantum wells in large magnetic fields
\cite{lilly1999, cooper02, du1999}.

Non-$s$-wave spin channel Pomeranchuk instabilities
are ``{\it unconventional magnetism}'' 
in analogy to unconventional superconductivity, which have been extensively
investigated\cite {hirsch1990, hirsch1990a, oganesyan2001,
wu2004, wu2007, varma2005, varma2006, kee2005}.
In Refs \cite{wu2004} by Wu and Zhang,
these states are classified as isotropic and anisotropic phases
dubbed $\beta$ and $\alpha$-phases as the counterparts of
$^3$He-$B$ (isotropic) and $A$ (anisotropic) phases, respectively.
The $\beta$-phases have circular or spherical Fermi surfaces
with topologically non-trivial spin configurations in momentum space
\cite{wu2004}.
In the $\beta$-phase, the relative spin-orbit symmetry is broken,
a concept introduced in the context of $^3$He-B phase,
while essentially the overall rotational symmetry is not.
The $\alpha$-phases are anisotropic electron liquid crystal states with spin
degree of freedom, which have been studied by many groups:
the $p$-wave phase was first studied by Hirsch 
\cite{hirsch1990, hirsch1990a} under the name of the 
``spin-split'' state, and was also proposed by Varma {\it et al.}
\cite{varma2005, varma2006}
as a candidate for the hidden order phenomenon in the heavy 
fermion compound URu$_2$Si$_2$;
the $d$-wave phase
was studied by Oganesyan {\it et al.} \cite{oganesyan2001} under the name of
``nematic-spin-nematic'' phase.
Systematic studies of the ground state properties  and collective 
excitations in both the anisotropic $\alpha$ and isotropic
$\beta$-phases has been performed
by Wu {\it et al.} \cite{wu2004, wu2007}.
Very recently, Chubukov and Maslov found that when approaching
the ferromagnetic quantum critical point, the $p$-wave channel
spin Pomeranchuk instability develops prior to the developing
of ferromagnetic instability \cite{chubukov2009}.

Although unconventional magnetism has not been convincingly identified
in experiments,
a spontaneous nematic electron liquid has been observed 
in the ultra-clean samples of the bilayer 
ruthenate Sr$_3$Ru$_2$O$_7$~\cite{grigera2001,grigera2004,borzi2007},
which has arouse much research interest \cite{berridge2008,millis2002,
green2004,tamai2008,iwaya2007,kee2005,yamase2007,puetter2007}.
Sr$_3$Ru$_2$O$_7$ is a metallic itinerant system with the tetragonal
RuO$_2$ $(ab)$ planes.
It is paramagnetic at zero magnetic field, and develops two consecutive 
metamagnetic transitions in the external magnetic field $B$ 
perpendicular to the $ab$-plane at 7.8 and 8.1 Tesla below 1K, respectively.
In the state between two metamagnetic transitions, the resistivity 
measurements show a strong spontaneous in-plane anisotropy (nematic) 
along the $a$ and $b$-axis with no noticeable lattice distortions, 
thus this effect is of electronic origin.   
It can be interpreted as a nematic state with the anisotropic 
distortion of the Fermi surface of the majority spin polarized by
the external magnetic field in Ref. \cite{grigera2004},
which is essentially a mixture of the $d$-wave Pomeranchuk instabilities 
in both density and spin channels.

In spite of years of intensive research, most theories remain 
phenomenological without considering the actual orbital band structures
of Sr$_3$Ru$_2$O$_7$ \cite{berridge2008,millis2002,green2004}.
Two key questions have not been 
answered satisfactorily after years of intensive research.
First, Sr$_2$Ru$_3$O$_7$ is a $t_{2g}$-band system, having both quasi-one
dimensional bands of $d_{xz}$ and $d_{yz}$ and the two-dimensional bands
of $d_{xy}$. Which are responsible for the nematic behavior?
Second, the nematic states require strong exchange interactions in the
$d$-wave channel, but the usual exchange interaction from Coulomb
repulsion is mostly in the $s$-wave channel.
Microscopic theories based on the single band picture of $d_{xy}$ 
combined with the van Hove singularity have been nicely 
developed \cite{kee2005,yamase2007,puetter2007}.
However, their models need a large $d$-wave channel exchange interaction
which was introduced by hand.
It is difficult to justify the microscopic origin of this
interaction in terms of Coulomb interaction.
Furthermore, these theories do not address the fact that 
the nematic ordering does not appear in
the single layer compound  Sr$_2$RuO$_4$, which has 
a similar band structure of $d_{xy}$.

In this article, we provide a natural answer to these two key questions
by extending the theory of Pomeranchuk instabilities into multi-orbital
systems.
We point out that it is the hybridized quasi-one-dimensional $d_{xz}$ 
and $d_{yz}$ bands instead of the $d_{xy}$ band that are responsible for 
the nematic ordering based on the following reasoning.
The key difference of electronic structures between Sr$_3$Ru$_2$O$_7$ and 
Sr$_2$RuO$_4$ is the bilayer splitting, which is prominent for the quasi-one
dimensional bands of $d_{xz}$ and $d_{yz}$ but small for the two-dimensional
bands of $d_{xy}$.
It is natural to expect that the spontaneous nematic behavior occurs in 
the bands of $d_{xz}$ and $d_{yz}$ and is accompanied by an orbital
ordering.
Furthermore, the orbital band hybridization between them
shifts a significant spectra weight of the exchange interaction 
into the $d$-wave channel, thus the nematic ordering can arise
from the conventional multi-band Hubbard interactions.
This mechanism also applies to other strongly correlated orbital
systems.

This paper is organized as follows.
We first present a heuristic picture to illustrate the idea 
how orbital hybridization enhances the Landau interaction in 
non-$s$-wave channels in Sect. \ref{sect:heur}.
In Sect. \ref{sect:GL}, a phenomenological Ginzburg-Landau
free energy is constructed to explain the two consecutive
metamagnetic transitions and the nematic phase in between.
In Sect. \ref{sect:micro}, we use a microscopic multi-orbital 
Hubbard model based on the quasi-one dimensional $d_{xz}$ and 
$d_{yz}$-bands to explain the appearance of the nematic state.
Conclusions and outlooks are made in Sect. \ref{sect:conclusion}.

\section{Landau interactions modified by orbital hybridization}
\label{sect:heur}
In this section, we present a heuristic picture to illustrate the enhancement 
of the non-$s$-wave Landau interactions from orbital 
band hybridizations.
For a single band system without orbital structures, 
the Landau interaction functions can be simply expressed at 
the Hartree-Fock  level in the density and spin channels as:
\bea
f^s(\vec p_1, \vec p_2)&=&V(\vec q=0)-\frac{1}{2} V(|\vec p_1-\vec p_2|),\nn\\
f^a(\vec p_1, \vec p_2)&=&-\frac{1}{2} V(|\vec p_1 -\vec p_2|),
\eea
where $V(\vec p)$ is the Fourier transform
of the two-body interaction $V(|\vec r_1-\vec r_2|)$, say, the Coulomb
interaction.
The high partial wave channel components of $V$ are usually weak, 
thus the condition for  Pomeranchuk instabilities in high 
partial channels is more stringent than that of the $s$-wave 
instability of ferromagnetism.

This situation is significantly changed in multi-band systems
with non-trivial orbital hybridization.
Let us consider a simplified two dimensional example of the hybridized 
bands between $d_{xz}$ and $d_{yz}$ and assume the single particle 
eigenstates around the Fermi surface takes the form of
\bea
|\psi_\sigma (\vec p)\rangle&=&e^{i\vec p \cdot \vec r} |u(\vec p)\rangle
\otimes \chi_\sigma, \nn\\
|u(\vec p)\rangle&=&\cos \phi_p |d_{xz}\rangle +\sin \phi_p |d_{yz}\rangle
\label{eq:orbit_cont}
\eea
where $\phi_p$ is the azimuthal angle of $\vec p$;
$|u(\vec p)\rangle$ is the Bloch wave function with the internal
orbital configurations; $\chi_\sigma (\sigma=
\uparrow, \downarrow)$ are the spin eigenstates.
This orbital structure has no effects on the Hartree interaction between 
two electrons with opposite spins, while it significantly changes the
Fock exchange interaction of two electrons of the same spin which
are not completely indistinguishable any more.
Consequentially, the exchange interaction between them
acquires an extra form factor describing the inner
product of their orbital configurations as 
\bea
-V(\vec p_1 -\vec p_2) |\avg{u(p_1)| u(p_2)}|^2.
\eea
The Landau interaction functions change to
\bea
f^s(\vec p_1,\vec p_2)&=& V(\vec q=0)
-\frac{1}{4}[1+\cos 2 
(\phi_{p_1}-\phi_{p_2})] \nn \\
&\times&
V(\vec p_1 -\vec p_2),
\nonumber \\
f^a(\vec p_1, \vec p_2)&=& -\frac{1}{4}[1+\cos 2 
(\phi_{p_1}-\phi_{p_2})] V(\vec p_1 -\vec p_2). \ \ \
\label{eq:landaufunc}
\eea

Therefore even if the bare interaction $V(\vec p_1-\vec p_2)$
is dominated by the $s$-wave channel, the extra $d$-wave
factor arising from the orbital hybridization
shifts a significant weight into the $d$-wave channel.

\section{The Ginzberg-Landau Theory}
\label{sect:GL}

The Eq. \ref{eq:landaufunc} above implies that a strong ferromagnetic 
or metamagnetic ($s$-wave) tendency also enhances the nematic 
ordering ($d$-wave).
Before constructing the microscopic theory, we illustrate this point 
through a Ginzburg-Landau free energy formalism which includes the 
coupling among the ferromagnetic
order $m$, the charge nematic order $n_c$, and spin nematic order $n_{sp}$.

\begin{figure}
\centering\epsfig{file=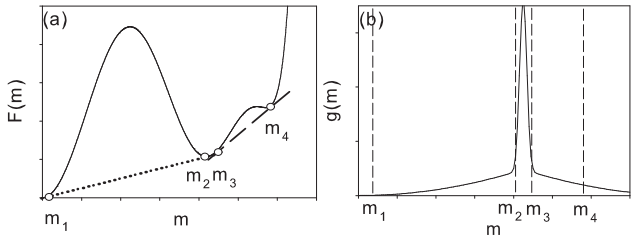,clip=1,width=\linewidth,angle=0}
\caption{a) The sketch of the magnetic part $F(m)$ in the
Ginzburg-Landau free energy. The slopes of two common tangent lines mark the
magnetic fields of meta-magnetic transitions.
b) The dimensionless coupling function $g'(m)$ has a peak distribution 
between two consecutive matamagnetic transitions.
}
\label{fig:GL}
\end{figure}

In the square lattice, the two different nematic channels 
$d_{x^2-y^2}$ and $d_{xy}$ belong to non-equivalent representations.
Only the $d_{x^2-y^2}$ channel instability is experimentally
observed and thus is kept.
Due to the experimentally observed anisotropy between the $z$-axis
and the $ab$-plane, we only keep the $z$-component of spin and spin-nematic
orders.
The Ginzburg-Landau free energy is constructed as
\bea 
F(h)&=& F(m) -h m 
+ r_c n_c^2 +r_{sp} n_{sp}^2 + g_c n_c^4 +g_s n_{sp}^4 \nn \\
&+& g(m) n_c n_{sp}
\label{eq:GL}
\eea
where 
$F(m)$ is the magnetic order contribution to the free energy
as an even function of $m$;
$h$ is the external magnetic field;
$r_c\propto(2+F^s_2)$ and $r_{sp}\propto (2+F^a_2)$ are the 
mass terms of charge and spin nematic orders, respectively;
$n_{c,sp}$ are defined as:
\bea
n_c&=&\sum_{\vec k, \sigma} \langle\psi^\dagger_\sigma(\vec{k})\psi_\sigma(\vec{k})
\rangle f(\vec k),\nn\\
n_{sp}&=&\sum_{\vec k, \sigma} \langle\psi^\dagger_\alpha(\vec{k})
\sigma_{z,\alpha\beta} \psi_\beta(\vec{k})\rangle f(\vec k),
\eea
where $f(\vec k)$ is a form factor exhibiting the $d_{x^2-y^2}$ symmetry;
$g(m)$ is the coupling function between $n_c$ and $n_{sp}$ which is an odd 
function of $m$ as required by time-reversal symmetry.
The experimentally observed two consecutive metamagnetic transitions
can be reproduced by a suitable designed form of $F(m)$ sketched in
Fig. \ref{fig:GL} $a$ where only the part with $m>0$ is shown.
It has two common tangent lines marked with dotted and dashed 
lines, which touch the curve at points of $m_{1,2}$
and $m_{3,4}$, respectively.
In the external magnetic field $h$, the solution of $m$ satisfies
the equation: 
\be
\frac{d}{dm} F(m) =h.
\ee
Therefore, the slopes of the two common tangent lines $h_{1,2}$ 
can be interpreted as the fields at which magnetization jumps
from $m_1$ to $m_2$ and from $m_3$ to $m_4$, i.e.,
metamagnetic transitions occur.
When $h$ lies between these two transitions,
the magnetization $m$ evolves continuously.

The development of nematic orders between the two successive metamagnetic
transitions is triggered by the $g(m)$-term.
Although the nematic instability is enhanced by Eq. \ref{eq:landaufunc},
they are still a weaker instability compared
to ferromagnetism (metamagnetism).
This is because the condition for  Pomeranchuk instability 
in the $d$-wave channel
in 2D, i.e., $F_{2}^{s,a}<-2$, is more stringent that in the
$s$-wave channel, i.e. $F_0^a<-1$.
We assume that the charge and spin nematic channels
are close to be critical but not yet,
i.e., $r_{c,sp}$ are small but positive.
Due to the hybridization term of $g(m)$, 
the eigen-order parameters arise from the diagonalization 
of the quadratic terms of $n_{c,sp}$ as:
\bea
n_+ &=&  \cos\theta n_c +\sin \theta n_s,\nn\\
n_- &=& -\sin\theta n_c +\cos \theta n_s,
\eea
where $\tanh 2\theta= 2g(m)/ (r_c-r_{sp})$.
The corresponding eigenvalues read:
\be
r_\pm = \frac{1}{2}\big\{r_c+r_{sp}\pm\sqrt{(r_c-r_{sp})^2+4 [g(m)]^2}\big\}.
\ee
The critical coupling for the $n_{+}$ channel to develop the instability is
\bea
[g'(m)]^2\equiv \frac{[g(m)]^2}{4 r_{c} r_{sp}}>1. 
\label{eq:condition}
\eea

In the presence of the van Hove singularity of DOS,
which is a common mechanism for meta-magnetism, all the parameters 
in the GL free energy could change significantly so that
the distribution of the dimensionless coupling function $g'(m)$
may not be smooth.
For the nematic order only occuring in the regime between two
metamagnetic transitions, $g'(m)$ must have a peak satisfying 
Eq. \ref{eq:condition} at $m_2<m<m_3$ but is below the critical 
value elsewhere as sketched in Fig. \ref{fig:GL} $b$.
Roughly speaking, the underlying physics is that 
metamagnetism pushes the majority Fermi surface
even closely to the van Hove singularity which finally drives the 
nematic ordering \cite{kee2005}.
The minority Fermi surfaces is pushed away from critical.
In the following we will confirm this mechanism explicitly in 
a microscopic calculation.

\begin{figure}
\centering\epsfig{file=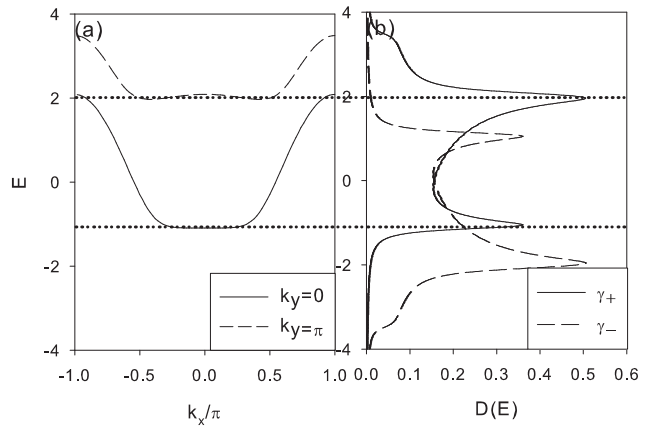,clip=1,width=\linewidth,angle=0}
\caption{\label{fig:dos} 
(a) Dispersion of the $\gamma_+$ band for $\vec{k}=(k_x,0)$ and $(k_x,\pi)$.
(b) The DOS of the $\gamma_\pm$ bands whose peaks corresponds to 
the van Hove singularities at $\vec{k}=(0,0), (0,\pi), (\pi,0)$.}
\end{figure}

\section{Microscopic Theory for the bilayer $Sr_3Ru_2O_7$} 
\label{sect:micro}
We next make the connection to Sr$_3$Ru$_2$O$_7$ by exploiting a microscopic 
model with $d_{xz}$ and $d_{yz}$-orbital bands.
They have the bonding and anti-bonding bands with a large
bilayer splitting at the order of their band widths.
Only the bonding bands, the even combination of the bilayer 
orbitals, are considered because their Fermi surfaces are close to van 
Hove singularity which enhances interaction effects.
Because of orbital hybridization, they form closed Fermi surfaces.
As moving around this Fermi surface, the orbital configuration 
varies between $d_{xz}$ and $d_{yz}$ exhibiting a $d$-wave pattern,
thus the mechanism illustrated in Sect. \ref{sect:heur} applies.

The band Hamiltonian $H_0$ reads
\bea
H_0&=&\sum_{\vec{k}\sigma} \big \{\epsilon_{xz,\vec{k}}d^\dagger_{xz\vec{k}\sigma} 
d_{xz,\vec{k}\sigma}+\epsilon_{yz,\vec{k}}d^\dagger_{yz,\vec{k}\sigma} 
d_{yz,\vec{k}\sigma}  \nn \\
&+& \lambda_{\vec{k}} (d^\dagger_{xz,\vec{k}\sigma} d_{yz, \vec{k}\sigma} + h.c.) \big\},
\eea
with
\bea
\epsilon_{xz,\vec{k}}&=&-2t_\pp\cos k_x-2t_\perp\cos k_y-4t'\cos k_x\cos k_y,\nn\\
\epsilon_{yz,\vec{k}}&=&-2t_\perp\cos k_x-2t_\pp\cos k_y-4t'\cos k_x\cos k_y,\nn\\
\lambda_{\vec{k}}&=&4t^{\prime\prime}\cos k_x\cos k_y.
\eea
$t_{\pp}$ and $t_{\perp}$ are the nearest neighbor longitudinal and transverse 
hopping integrals for the $d_{xz}$ and $d_{yz}$-orbitals;
$t^\prime$ and $t^{\prime\prime}$ are the next-nearest neighbor 
intra and inter-orbital hopping, respectively.
The resulting diagonalized band Hamiltonian is:
\be
H_0=\sum_{\vec{k}\sigma} E^+_{\vec{k}}\gamma^\dagger_{+\vec{k}\sigma}
\gamma_{+\vec{k}\sigma} + E^-_{\vec{k}}\gamma^\dagger_{-\vec{k}\sigma}
\gamma_{-\vec{k}\sigma},
\ee
where
\bea
E^\pm_{\vec{k}}&=&\frac{1}{2}\left[\epsilon_{xz,\vec{k}}+\epsilon_{yz,\vec{k}}
\pm\sqrt{(\epsilon_{xz,\vec{k}}-\epsilon_{yz,\vec{k}})^2+4h^2_{\vec{k}}} \right]
\eea
and the band eigen-operators reads
\bea
\gamma_{+, \vec{k}\sigma}&=&\cos \phi_{\vec{k}} d_{xz,\vec{k}\sigma} 
+ \sin \phi_{\vec{k}} d_{yz,\vec{k}\sigma},\nn\\ 
\gamma_{-,\vec{k}\sigma}&=&-\sin \phi_{\vec{k}} d_{xz,\vec{k}\sigma}
 + \cos \phi_{\vec{k}} d_{yz,\vec{k}\sigma} 
\eea
with the hybridization angle $\phi_{\vec k}$ satisfying
\bea
\tan 2 \phi_{\vec k}= \frac{2 \lambda_{\vec k}}
{\epsilon_{xz,\vec{k}}-\epsilon_{yz,\vec{k}}}
=\frac{-4t^{\prime\prime}}{t_\pp-t_\perp}
\frac{\cos k_x \cos k_y}{\cos k_x - \cos k_y}.
\label{eq:angle}
\eea
This band hybridization pattern is a lattice version of Eq. 
\ref{eq:orbit_cont} with only the $d_{x^2-y^2}$ channel but
not the $d_{xy}$ channel.
We choose the parameter values of $(t_\pp,t_\perp,t',t^{\prime\prime})
=(1.0,0.145,0.0,0.3)$, and plot the density of states (DOS) 
in Fig \ref{fig:dos}.
Two peaks in the DOS exist in both $\gamma_{\pm}$ bands,
which correspond to the van Hove singularities at 
$\vec{k}=(0,0), (0,\pi), (\pi,0)$.

We take the general multi-band Hubbard model 
that are widely used in literatures for the interactions as
\bea
H_{int}&=&U\sum_{i,a=xz,yz} n_{a\uparrow}(i)n_{a\downarrow}(i) 
+V\sum_i n_{xz}(i)n_{yz}(i) \nn \\
&-&J\sum_i \{ \vec{S}_{xz}(i)\cdot\vec{S}_{yz}(i)-\frac{1}{4}n_{xz}(i)n_{yz}(i)\}
\nn\\
&+& \Delta \sum_i \{ d^\dagger_{xz,\uparrow}(i) d^\dagger_{xz,\downarrow}(i)
d_{yz,\downarrow} (i) d_{yz,\uparrow}(i) +h.c.\}, \nn \\
\label{eq:hubbard}
\eea
where $n_{a,\sigma}$ are particle number operators in orbital $a$
with spin $\sigma$; $n_a=n_{a,\uparrow}+n_{a,\downarrow}$;
$\vec{S}_a$ are spin operators in orbital $a$.
The $U$-term is the intra-orbital repulsion; 
the $V$-term is the inter-orbital repulsion for the spin triplet
configuration of two electrons;
the $J$-term represents the Hund's rule physics;
the $\Delta$-term describes the inter-orbital pairing hopping.

In the absence of the orbital hybridization, it is not conclusive that 
the Hubbard model can give rise to nematic transitions because
on-site interactions only contribute to the $s$-wave channel.
We define
the charge nematic order $2n_c=n_{xz}-n_{yz}$, the 
spin nematic order $n_{sp}=S^z_{xz}-S^z_{yz}$, and 
the ferromagnetic order $m=S^z_{xz}+S^z_{yz}$.
The mean-field theory in the eigen-basis of $\gamma_{\pm}$ reads:
\be
H_{mf}=\sum_{\vec{k}\sigma,\alpha=\pm} \xi_{\alpha\sigma}
\gamma^\dagger_{\alpha\vec{k}\sigma}\gamma_{\alpha\vec{k}\sigma} 
+ V_m\,m^2 + V_{sp}\,n_{sp}^2 + V_c n_c^2,
\ee
where 
\bea
V_m=\frac{U}{2}+\frac{J}{4}, ~ V_{sp}=\frac{U}{2}-\frac{J}{4}, ~
V_c=V+\frac{J}{4}-\frac{U}{2},
\eea
and the less important inter-band coupling terms
$\gamma^\dagger_{+\vec{k}\sigma}\gamma_{-\vec{k}\sigma}$ are dropped.
The renormalized single particle spectra become:
\be
\xi_{\alpha\sigma}=E^\alpha_{\vec{k}} - \mu - \sigma V_m m - 
\alpha (V_c n_c + \sigma V_{sp}n_{sp} )\cos 2\phi_k,
\ee
where $\phi_k$ are the hybridization angle defined in Eq. \ref{eq:angle}.
The self-consistent equations for the order parameters read:
\bea
n_c&=&\frac{1}{2N}\sum_{k,\sigma} 
\{n_{f,+,\sigma}(\vec k)-n_{f,-,\sigma}(\vec k) \} \cos 2\phi_{\vec k}, \nn\\
n_{sp}&=&\frac{1}{2 N}\sum_{\vec{k},\sigma}\sigma 
\{ n_{f,+,\sigma}(\vec k)-n_{f,-,\sigma}(\vec k)\}\cos 2\phi_{\vec{k}}, \nn\\
m&=&\frac{1}{2 N}\sum_{\vec{k},\sigma}
\sigma \{ n_{f,+,\sigma}(\vec k)+n_{f,-,\sigma}(\vec k) \},
\eea
where the $\cos 2 \phi_{\vec k}$ factor represents the
$d_{x^2-y^2}$ symmetry of the nematic orders of $n_c$ and $n_{sp}$.

\begin{figure}
\centering\epsfig{file=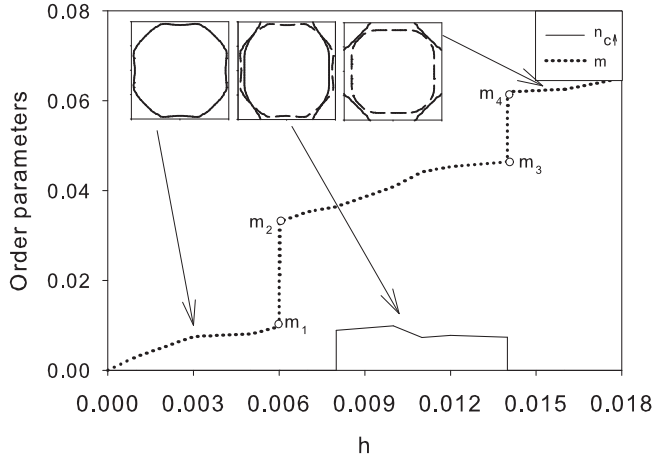,clip=1,width=\linewidth,angle=0}
\caption{\label{fig:order} 
$m$ and the nematic order parameter for majority spin band 
($n_{c\uparrow}=(n_c+n_{sp})/2$) {\it versus} the external field $h$. 
The topologies of the FS of the $\gamma_+$ band in each phases are
sketched in the insets, where the solid and dashed lines represent 
the FS of the majority and the minority spin bands. 
}
\end{figure}

We are ready to study the zero-temperature phase diagram with the
Zeeman energy term as:
\be
H_{ext}= -h\sum_{a=xz,yz}  \sigma n_{a,\sigma}(i).
\ee
$\mu$ is tuned to reach the filling of $n=3.48$ at $h=0$ in the bilayer bonding
bands of $d_{xz}$ and $d_{yz}$ so that the FS of the $\gamma_+$ band is very 
close to the van Hove singularity while that of the $\gamma_-$ band 
is far away from the singularity.
The interaction parameters are chosen as $(U,V,J,\Delta)=(2.7,2.5,0,0)$,
which remarkably reproduce the isotropic-nematic-isotropic phase transition 
as a function of $h$ as we expect from the Ginzburg-Landau analysis. 

The magnetization $m$ and the nematic order of the majority spin Fermi surface 
$n_{c,\uparrow}=\frac{1}{2}(n_c + n_{sp})$
are depicted in Fig. \ref{fig:order}, while that 
of the minority spin Fermi surface is much smaller 
(not shown) because its Fermi surface
is pushed from the van Hove singularity.
As explained in Ref. \cite{kee2005}, the first jump of $m$ distorts
the Fermi surface to touch the van Hove singularity points along one of the $ab$
axes, and the second jump restores the four fold symmetry of Fermi
surface to cover the singularity in both directions.
Compared to Ref. \cite{kee2005}, the range of $h$ for the nematic 
state is significantly reduced in agreement with experiment due
to inclusion of the magnetic order $m$ self-consistently in the
solution.
Furthermore, the first jump of $m$ is larger than the second 
one as consistent with the experiments. 
This feature can be understood by the asymmetric profile of the 
DOS at the van Hove singularity which drops faster in the higher
energy side as shown in Fig. \ref{fig:dos}.
After the first transition, part of the FS of the majority spin
has moved beyond the van Hove singularity.
This reduces the DOS at the second jump and leads to a weaker 
second jump where the nematic order disappears.

We learned another independent and beautiful work by Raghu {\it et al.} 
\cite{raghu2009} which was posted at the same time as ours.
The same mechanism is proposed for the nematic state observed in 
Sr$_2$Ru$_3$O$_7$ based on the quasi-one dimensional bands.
They used a more detailed band structure and further considered
the spin-orbit coupling effect.
We further performed the Ginzburg-Landau analysis for the
competition between ferromagnetization and nematic ordering.

\section{Conclusion and Outlook}
\label{sect:conclusion}

In this article, we have studied the Fermi liquid 
instability of the Pomeranchuk type in orbital band
systems.
Orbital band hybridization significantly enhances the Landau 
interaction functions in high partial wave channels, thus providing 
a mechanism for the nematic states or unconventional magnetism
from conventional interactions.
Consequentially, metamagnetism (ferromagnetism) induces 
the nematic behavior even with the onsite multi-band 
Hubbard interactions.

This mechanism is applied to the $t_{2g}$ system of
Sr$_2$Ru$_3$O$_7$ by attributing the observed nematic behavior to 
the hybridized quasi-one dimensional bands of $d_{xz}$ and $d_{yz}$,
which is the major difference between our work and Ref.
\cite{kee2005,puetter2007,yamase2007}.
Many open questions still need future exploration.
In particular, the quick suppression of the nematic behavior
with the in-plane magnetic field $h_\pp$ might result from
the orbital effect due to the bilayer splitting as pointed out
in Ref. \cite{puetter2007}, or from the spin-orbit coupling
effect.

The mechanism presented in this article also very general.
It is essentially a Berry phase effect which naturally
arises from electron systems with non-trivial band structures,
such as spin-orbit coupling system and graphene.
It can be understood as a conventional interaction acquires
a non-trivial nature after projected onto a non-trivial band 
structure.
We further predict that the nematic ordering arises at the ferromagnetic 
transition in spin-orbit coupling systems
as a result of the hybridization between 
two spin components in the band structure.
For example, the in-plane ferromagnetic ordering in the quasi-2D
Rashba systems and the surface states of the topological 
insulators induces the $p$-wave distortion of Fermi surfaces,
which will be presented in a later publication \cite{lee2008}.

\section{Acknowledgement}
We thank X. Dai, J. Hirsch, Y. B. Kim and H. Y. Kee for helpful discussions.
C. W. is grateful for  E. Fradkin, S. A. Kivelson, 
and S. C. Zhang for their education on Pomeranchuk instability
and encouragement on this project.
This work is supported by ARO-W911NF0810291
and Sloan Research Foundation.



\end{document}